\documentclass[showpacs,preprintnumbers,amsmath,amssymb,prd, nofootinbib]{revtex4}
\usepackage{amsmath,amssymb,graphicx}
\usepackage{epsf}
\usepackage{epstopdf}
\usepackage {amssymb}
\usepackage {verbatim}
\newcommand{\nc}{\newcommand}
\nc{\ba}{\begin{eqnarray}}
\nc{\ea}{\end{eqnarray}}
\newcommand\be{\begin{equation}}
\newcommand\ee{\end{equation}}

\newcommand\mPl{{M_{\rm Pl}}}

\newcommand{\measure}[1]{{\mathrm{d}{{#1}}}}
\newcommand{\dn}[2]{{\mathrm{d}^{{#1}}{{#2}}}}	

\newcommand{\calR}{{\cal{R}}}
\newcommand{\calH}{{\cal{H}}}
\nc{\x}{{\bf{x}}}

\begin{document}

\title{  Fluid Inflation}

\author{Xingang Chen$^{1,2}$}
\author{Hassan Firouzjahi$^{3}$}
\author{Mohammad Hossein Namjoo$^{4}$}
\author{Misao Sasaki$^{5}$}
\affiliation{$^1$Centre for Theoretical Cosmology, DAMTP, University of Cambridge, Cambridge CB3 0WA, UK}
\affiliation{$^2$ Department of Physics, The University of Texas at Dallas, Richardson, TX 75083, USA}
\affiliation{$^3$School of Astronomy, Institute for Research in
Fundamental Sciences (IPM),
P. O. Box 19395-5531,
Tehran, Iran}
\affiliation{$^4$School of Physics, Institute for Research in
Fundamental Sciences (IPM),
P. O. Box 19395-5531,
Tehran, Iran}
\affiliation{$^5$Yukawa Institute for theoretical Physics,
 Kyoto University, Kyoto 606-8502, Japan}

\date{\today}

\begin{abstract}
\vspace{0.3cm}
In this work we present an inflationary mechanism based on fluid dynamics.
Starting with the action for a single barotropic perfect fluid,
we outline the procedure to calculate the power spectrum and the
bispectrum of the curvature perturbation.
It is shown that a perfect barotropic fluid naturally gives rise to
a non-attractor inflationary universe in which the curvature perturbation
is not frozen on super-horizon scales.
We show that a scale-invariant power spectrum can be obtained with
the local non-Gaussianity parameter $f_{NL}= 5/2$.

\vspace{0.3cm}

\end{abstract}

\preprint{YITP-13-41, IPM/A-2013/25}

\maketitle

\section{Introduction}

Cosmic inflation has emerged as a very successful paradigm for
the early universe and structure formations. The basic predictions of
simple models of inflation for the curvature perturbation power spectrum
and bispectrum are in very good agreement with recent cosmological observations
 such as WMAP~\cite{Hinshaw:2012fq} and PLANCK~\cite{Ade:2013lta, Ade:2013uln}.

Without fully addressing the UV completion
aspects of inflation, at the low energy effective field theory
level, one can explore a variety of possibilities in the
inflationary model building. In fact, many models of inflation based
on scalar fields are constructed purely phenomenologically.
Furthermore, one may add various features to such models,
 for example, by introducing extra phenomena such as particle creation
and field annihilation, or local departures from inflation
such as steps in the potential, turning trajectories and waterfall mechanisms.
These additions have been used to explain
the local features or glitches seen in CMB
observations \cite{Starobinsky:1992ts, Leach:2001zf, Adams:2001vc, Gong:2005jr, Chen:2006xjb, Joy:2007na, Hotchkiss:2009pj, Abolhasani:2012px, Abolhasani:2010kn,
Arroja:2011yu, Adshead:2011jq, Miranda:2012rm, Adshead:2013zfa, Achucarro:2012sm, Achucarro:2012fd, Cremonini:2010ua, Avgoustidis:2012yc,  Gao:2012uq, Romano:2008rr, Ashoorioon:2006wc, Battefeld:2010rf, Firouzjahi:2010ga, Battefeld:2010vr, Barnaby:2009dd, Barnaby:2010ke, Biswas:2010si, Chen:2008wn, Bean:2008na, Silverstein:2008sg, Flauger:2009ab, Flauger:2010ja, Chen:2010bka, Chen:2011zf, Chen:2011tu, Battefeld:2013xka, Saito:2013aqa, Namjoo:2012xs}.

In this work we consider a different type of low energy effective
field theory model for inflation. Namely, we present a formalism to obtain
inflation from a fluid. Our starting point is the Lagrangian formalism
for a perfect fluid in Einstein gravity,
which enables us to calculate the power spectrum and bispectrum
of the curvature perturbation.

Depending on the equation of state and whether it is an isentropic
(barotropic) or non-isentropic fluid, different inflationary scenarios
are possible. As a first step, we concentrate on an isentropic fluid
in which the pressure is a given function of the energy density.
In principle, one should be able to extend this formalism to a
non-isentropic fluid.

This paper is organized as follows. In Section \ref{formalism} we present
a Lagrangian formalism for fluid inflation and the background equations.
In Section \ref{perturbations} we present the cosmological perturbation
theory in our setup and calculate the power spectrum and bispectrum of
the curvature perturbation.
In Section \ref{Model} we present a simple scalar field model
that mimics our fluid model. We then coclude the paper with
a short discussion.

\section{The Formalism}
\label{formalism}

To calculate the power spectrum and the bispectrum we need to have a
Lagrangian formalism of fluid dynamics coupled with Einstein gravity.
Here we use the Lagrangian for the perfect fluid in the presence of gravity
 proposed by Ray \cite{Ray1, Ray2}
\ba
{\cal L}= \dfrac{1}{2} \mPl^2 \sqrt{-g} R - \sqrt{-g} \, {\rho} (1+ e(\rho)) + \sqrt{-g} \lambda_1 \left(  g_{\mu \nu} U^{\mu} U^\nu +1 \right) + \sqrt{-g} \, \lambda_2 \left( \rho U^\mu   \right)_{;\mu}\,,
\ea
where $\rho$ is the rest mass density, $e(\rho)$ is the
specific internal energy, $U^{\mu}$ is the 4-velocity
and $\lambda_1$ and $\lambda_2$ are Lagrange multipliers for
the two constraints; the first is the normalization of
the 4-velocity and the second is the conservation of the rest mass density.
Note that the total energy density, $E$, is given by
\ba
\label{E-eq}
E = \rho (1+e)\,.
\ea
Below we show that the above Lagrangian gives the correct equations
of motion for an isentropic perfect fluid minimally coupled to gravity.
In this work we concentrate on an isentropic fluid for which $e= e(\rho)$.
In principle one can consider more general situations in which $e$
is als a function of other thermodynamic variables such as entropy.

Varying the action with respect to the Lagrange multipliers
$\lambda_1$ and $\lambda_2$ yields
the normalization condition for $U^\mu$ and the energy conservation
equations, respectively,
\ba
\label{u-normal}
U^\mu U_\mu = -1\,,
\ea
and
\ba
\label{energy}
\left( \rho U^\mu \right)_{; \mu} =0 \,.
\ea

Varying the action with respect to $\rho$ and $U^{\mu}$
yields, respectively,
\ba
\label{C1}
\lambda_{2,\mu } \, U^\mu = - \dfrac{d E}{d \rho}\,,
\ea
\ba
\label{C2}
 \lambda_1 = \dfrac{1}{2} \rho  \dfrac{d E}{d \rho}\,,
\ea
where the constraint Eq.~(\ref{u-normal}) have been used
to obtain the latter equation.

Finally, varying the action with respect to $g_{\mu \nu}$ yields
the Einstein equation,
\ba
G^{\mu \nu} = \dfrac{1}{\mPl^2} T^{\mu \nu}  \, ,
\ea
where $G^{\mu \nu}$ is the Einstein tensor and the
energy momentum tensor $T^{\mu\nu}$ is given by
\ba
T^{\mu \nu} = \rho  \dfrac{d E}{d \rho}  U^\mu U^\nu
 + g^{\mu \nu} \left( \rho  \dfrac{d E}{d \rho} -E \right)\,.
\ea

In addition to the above Euler-Lagrange equations,
using the second law of thermodynamics, one has
\ba
T ds = de + P d (\dfrac{1}{\rho})\,,
\ea
where $s$ is the entropy density and $P$ is the pressure.
For an isentropic fluid we have $ds=0$, hence
\ba
\label{P}
\dfrac{d e (\rho)}{d\rho}= \dfrac{P}{\rho^2}\,.
\ea
Knowing that $e=e(\rho)$, the above equation also implies that $P$
is a function of $\rho$. Alternatively, in terms of the energy density $E$,
using Eq.~(\ref{E-eq}) we obtain
\ba
\label{E-diff}
\frac{d E}{d \rho} = \frac{E+P}{\rho} \,.
\ea
Equations (\ref{P}) and (\ref{E-diff}) imply that $P$ is a function of $E$,
$P=P(E)$, which is expected for an isentropic or barotropic fluid.
Plugging Eq.~(\ref{E-diff}) into the definition of $T^{\mu \nu}$ yields
\ba
T^{\mu \nu}=(E + P) U^\mu U^\nu+ P g^{\mu \nu}\,.
\ea
Thus we recover the standard form for the energy momentum tensor
of a perfect fluid.
\subsection{The background equations}

Here we provide the background equations.  As for the background, we assume a
flat FLRW universe,
\ba
ds^2=-dt^2 + a(t)^2 d \x^2\,.
\ea
Noting that at the background level $U^\mu= (1,0,0,0)$, from
Eqs.~(\ref{C1}) and (\ref{C2})
one obtains the equations for the Lagrange multipliers as
\ba
\lambda_1 = \dfrac{1}{2} (E + P)\,,
\qquad  \dot \lambda_2 = -\dfrac{1}{\rho} (E + P)\,.
\ea
Furthermore, the rest mass conservation equation (\ref{energy}) yields
\ba
\label{energy2}
\dot \rho + 3 H \rho = 0\,,
\ea
where $H \equiv \dot a/ a$ is the Hubble expansion rate.
The background Einstein equations are
\ba
\label{Friedmann}
\left( \frac{\dot a}{a} \right)^2 &=& \frac{E}{ 3 M_P^2}\,,
 \\
\frac{\ddot a}{a} &=& -\frac{E + 3 P}{6 M_P^2} \, .
\ea
Combining the above Einstein equations, one can easily recover the
energy conservation equation in an expanding background,
\ba
\label{back-cons}
\dot E + 3 H (E + P) = 0 \,.
\ea

Now we consider the inflationary background. First, let us look at the slow-roll
parameter $\epsilon \equiv - \dot H/H^2$.
Using the background Friedmann equation (\ref{Friedmann})
 and the energy conservation equation (\ref{back-cons}),
one has
\ba
\epsilon = -\frac{\dot H}{H^2} = \frac{E +P}{2 M_P^2 H^2}\,.
\ea
The second slow-roll parameter $\eta$ is given by
\ba
\label{eta-eq}
\eta \equiv \frac{\dot \epsilon}{H \epsilon} = 2 \epsilon - 3 (1 + c_s^2) \, .
\ea
Here the speed of sound $c_s$ for our isentropic fluid is given by
\ba
\label{cs-def}
c_s^2 \equiv \frac{\dot P}{\dot E}\,.
\ea
For an infinitesimal perturbation, this implies
\ba
\label{P-rho}
\delta P = c_s^2 \delta E = c_s^2 (E+ P) \frac{\delta \rho}{\rho}\,.
\ea
Note that the definition (\ref{cs-def}) is relevant
since we consider an isentropic fluid.

It is important to note that for stable perturbations with $c_s^2 >0$,
the magnitute of the $\eta$ parameter is never smaller than unity
as clear from Eq.~(\ref{eta-eq}).
Indeed, taking $\epsilon \ll 1$ to sustain a long enough period of inflation,
one obtains $\eta \simeq -3(1+ c_s^2)$. As we shall see below,
to have an almost scale-invariant power spectrum, we must require
$c_s \simeq 1$. So we conclude $\eta \simeq -6$.
This signals that our fluid inflationary system is within the domain of
``ultra slow-roll inflation''
scenarios~\cite{Kinney:2005vj, Namjoo:2012aa, Martin:2012pe, Huang:2013oya, Chen:2013aj}.
For a nearly constant $\eta$, one obtains
\ba
\label{epsilon-eq}
\epsilon( t) = \epsilon_i \left( \frac{a(t)}{a_i} \right)^{\eta}\,,
\ea
where $\epsilon_i$ is the value of $\epsilon$ at an initial/reference
time $t=t_i$. The fact that $\eta \simeq -6$ as explained above implies
that $\epsilon$ decays during the ultra slow-roll inflation like $a^{-6}$.

It is also instructive to look at the equation of state parameter $w\equiv P/E$.
Using the relation  $\dot P = c_s^2 \dot E$ and the background Friedmann
and the energy conservation equations,
one can easily check that
\ba
\label{w-eq}
\dot w =  - 3H (1 + w) (c_s^2 - w)\,.
\ea
We are interested in model in which the fluid has a (nearly) constant
sound speed. With a constant $c_s$, the above equation can be integrated,
 yielding
\ba
\label{w-eq2}
w = - \frac{1- F c_s^2}{1+ F}    \qquad , \qquad F \equiv \frac{1+ w_i}{c_s^2 - w_i} e^{- 3 N (1 +c_s^2)}  \, ,
\ea
where $w_i$ is the initial value of $w$.
As inflation proceeds, $F$ rapidly decays and one has
\ba
1+w \simeq (1+ c_s^2) F \propto e^{- 3 N (1 +c_s^2)} \, .
\ea
This means that $w$ approaches $-1$ exponentially rapidly.
As mentioned before, this means we are within the domain of
ultra slow-roll inflation.

Finally, with the assumption that $w \simeq -1$ and $\epsilon$ is
rapidly decaying, the background can be approximated by a pure
de Sitter solution to a high accuracy,
\ba
\calH(\tau) = \dfrac{\calH_e}{1+\calH_e (\tau_e-\tau)}\,,
\qquad a(\tau) = \dfrac{a_e}{\calH_e(\tau_e-\tau)+1}\,.
\ea
Here $\tau$ is the conformal time, $d \tau = dt/a(t) $,
$\calH=a'/a$ is the conformal Hubble parameter,
and the subscript $e$ denotes the value of a quantity
at the end of ultra slow-roll inflation.

\section{The perturbation}
\label{perturbations}

Now we consider the perturbation in our fluid coupled to gravity. For relevant studies in different context see \cite{Dubovsky:2005xd, Endlich:2012pz}.  For this purpose, we employ the ADM formalism in which the
metric components are expressed as
\begin{equation}
  \label{vertex:adm}
  \measure{s}^2 = - N^2 \measure{t}^2 + h_{ij} (\measure{x}^i + N^i
  \measure{t})(\measure{x}^j + N^j \measure{t})\,.
\end{equation}
Plugging the above into the action yields
\begin{eqnarray}
\label{vertex:action}
S &=& \int \measure{t} \, \dn{3}{x}\;\sqrt{h} N \left(L_G + L_m\right)\,;
\cr
\cr
&&
L_G =\frac{\mPl^2}{2}
\left[R^{(3)}+ N^{-2} (K_{ij} K^{ij} - K^2)\right]\,,
\end{eqnarray}
where $L_G$ is the gravitational part of the Langrangian,
$K_{ij}$ is the extrinsic curvature of the $t=\mathrm{constant}$
hypersurface,
\begin{equation}
  K_{ij}
= \frac{1}{2} \dot{h}_{ij} - ^{(3)}\nabla_j N_{i}-  ^{(3)}\nabla_i N_{j} \, ,
\end{equation}
in which $^{(3)}\nabla$ represents the covariant derivative with respect to
the three-dimensional metric $h_{ij}$ and $K$ is the trace of $K_{ij}$.
The matter Lagrangian $L_m$ is given by
\ba
\label{Lm}
L_m=-  \, {\rho} \left(1+ e(\rho) \right)
+ \lambda_1 \left(  g_{\mu \nu} U^{\mu} U^\nu +1 \right)
+  \, \lambda_2 \left( \rho U^\mu   \right)_{;\mu}\,.
\ea
Note that, by integration by parts, the above Lagrangian
density is equivalent to Lagrangian density $\hat L_m$
\ba
 \hat L_m=-  \, {\rho} (1+ e(\rho))
+ \lambda_1 \left(  g_{\mu \nu} U^{\mu} U^\nu +1 \right)
-  \, \lambda_{2;\mu} \left( \rho U^\mu   \right)\,.
\ea

The lapse function $N$ and the shift vector $N_i$ are Lagrange
multipliers. Varying the action with respect to them gives
the Hamiltonian and momentum constraint equations,
\ba
\label{C3}
  \mPl^2 R^{(3)} + 2{ L}_m
+ 2 N \dfrac{\partial { L}_m}{\partial N}
 - \frac{\mPl^2}{N^2} (K_{ij} K^{ij} - K^2) = 0\,,
  \\
\label{C4}
\mPl^2  \left[ \frac{1}{N} (K^{ij} - K h^{ij}) \right]_{;j}
 + N \dfrac{\partial { L}_m}{\partial N_i} = 0  \, .
\ea

\subsection{Linear perturbation}
Now we consider linear perturbations.
To proceed further, we have to choose a gauge. Since our system is based on
the fluid dynamics, it is convenient to choose the comoving gauge in
which\footnote{The terminology ``comoving gauge" used here
is somewhat different from the standard definition of the comoving gauge.
As seen from its definition (\ref{comoving}), here it is defined by
a time-slicing in which the fluid 4-velocity is orthogonal
to $t=const.$ hypersurfaces and the 3-metric is conformally flat.}
\ba
\label{comoving}
  U_{\mu} = (-1+u , 0,0,0)\,,
 \quad  h_{ij} = a^2(t) e^{2\calR } \delta_{ij} \, .
\ea
Here $u$ represents the velocity scalar potential to all order
in perturbations and $\cal R$ denotes the curvature perturbations
in the comoving gauge.

As usual we decompose the lapse and the shift functions into its scalar
 degrees of freedom,
\ba
N_i = \partial_i \psi\,, \qquad N= 1+\alpha\,.
\ea
Similarly, we perturb the Lagrange multipliers $\lambda_i$
and the density field $\rho$ as
\ba
 \lambda_i = \lambda_i^0 + \delta \lambda_i\,,
\qquad \rho = \rho^0 + \delta \rho \, .
\ea
In the above decompositions, we have focused on the scalar perturbations
and neglected  the tensor and vector perturbations.
From now on we omit the superscript $0$ from the background quantities.

Now we obtain the perturbed field equations. Perturbing the normalization
condition~(\ref{u-normal}) and the rest mass conservation
equation~(\ref{energy}) yields
\ba
\label{u-alpha}
\alpha+ u &=& 0\,,  \\
\label{energy-pert}
\dot{\delta \rho} + 3 H \delta \rho + 3 \rho \dot \calR
  - \rho \dfrac{\nabla^2}{a^2} \psi &=& 0\,.
\ea
Perturbing the expressions for the Lagrange multipliers $\lambda_i$ in
Eqs.~(\ref{C1}) and (\ref{C2}) yields
\ba
\label{delta-lambda1}
\delta \lambda_1 &=& \dfrac{1}{2} \delta P + \dfrac{\delta \rho}{2 \rho} (E+P)\,,
\\
\label{delta-lambda2}
\dot{\delta \lambda}_2 &=& -\dfrac{E+P}{\rho} \alpha - \dfrac{\delta P}{\rho}\,.
\ea
Furthermore, perturbing the constraint equations \eqref{C3} and \eqref{C4}
results in
\ba
\label{deltalambda2}
\dfrac{\nabla^2}{a^2} (\calR + H \psi) + 3 H (H \alpha - \dot \calR)
 + \dfrac{\delta \rho}{2 \rho \mPl^2} (E+P) = 0\,,
\\
 H \alpha  -  \dot \calR + \dfrac{ \rho \delta \lambda_2}{2\mPl^2} = 0\,.
\ea

Alternatively, one can perturb the Einstein equations.
In particular, the ($0i$)-component of the Eistein equations gives
\ba
\dot \calR = \alpha H\,.
\ea
Comparing this equation with \eqref{deltalambda2} yields $\delta \lambda_2=0$.

The other components of the Einstein equations are not necessary thanks
to the contracted Bianchi identities, or the energy momentum conservation
law $T^\mu{}_{\nu\,;\mu}=0$.
From the momentum conservation equation, $T^{\mu}{}_{\, i\, ;\mu}=0$,
one has
\ba
\delta P &=& - (E+P) \alpha\,.
\ea
Again this is consistent with the constraint (\ref{delta-lambda2})
if $\delta \lambda_2 = 0$.
Perturbing the energy conservation equation, $T^{\mu}{}_{\, 0\, ;\mu}=0$,
gives
\ba
\dot{\delta E} + 3 H \delta E + 3(E+P) \dot \calR
+ 3 H \delta P - (E+P) \dfrac{\nabla^2}{a^2} \psi=0  \, .
\ea
This equation can be obtained using the constraint equations
as well as the relation between $\rho$, $E$ and $P$, mentioned before.

By setting $\delta \lambda_2=0$ in the constraint equations and solving
for all the variables but $\calR$, one obtains an equation of motion
for $\calR$ which represents the unique propagating degree of freedom,
\ba
\label{R-eq}
\dfrac{\nabla^2}{a^2} \calR + 3 H \dot \calR
- H^2 \left(\dfrac{\dot \calR }{c_s^2 H^2} \right)^. =0 \, ,
\ea
where we recall that the sound speed $c_s$ is defined in Eq.~(\ref{cs-def}),
and it appears in the perturbed relations (\ref{P-rho}),
namely,
\ba
\delta P = c_s^2 \delta E = c_s^2 (E+ P) \frac{\delta \rho}{\rho}\,.
\ea

\subsection{Power spectrum}
\label{Power}
To calculate the power spectrum we need to expand the action to second order.
Let us firt recapitulate the action given by Eq.~(\ref{vertex:action}),
\ba
\label{S-general}
S = \int d^4 x \left[ \mPl^2 {\cal L}_{G} + N \sqrt{h} L_m \right]\,,
\ea
where ${\cal L}_G=N\sqrt{h}\,L_G$.
Accordingly the second order action is given in the form,
\ba
S_2 = \int d^4 x \left[ \mPl^2 {\cal L}_G^{(2)}
+ a^3 \left( { L}_m^{(2)} + (\alpha+3 \calR) { L}_m^{(1)}
+ (3 \alpha \calR +\dfrac{9\calR^2 }{2}) { L}_m^{(0)} \right) \right]\,,
\ea
where we have
\ba
L_m^{(0)} &=& - \rho (1+e) - \dot{\lambda}_2 \rho = P \,,
\nonumber \\
L_m^{(1)} &=& - \delta \rho (1+e) - \dfrac{P}{\rho} \delta \rho
+ \lambda_1 (2 \alpha +2 u) - \dot{\lambda}_2 \delta \rho
- \dot{\delta \lambda}_2 \rho + \dot \lambda_2 \rho (2 \alpha + u)
\\ \nonumber
&&= 2 \alpha (\lambda_1 + \rho \dot  \lambda_2 ) - \rho  \dot{\delta \lambda_2}\,,
\\
L_m^{(2)} &=& - \dfrac{1}{2 \rho} \dfrac{d P}{d \rho} \delta \rho^2
 - \lambda_1 (3 \alpha^2 + u^2) + \delta \lambda_1 (2 \alpha + 2 u)
- 3 \dot \lambda_2  \rho \alpha^2  + \dot \lambda_2 \delta \rho ( u + 2 \alpha )
\\ \nonumber
&&- \dot{\delta \lambda}_2 \left[\delta \rho - \rho (u+2 \alpha) \right]
+ \delta \lambda_{2,i} \rho \dfrac{\psi_{,i}}{a^2}\,,
\ea
and
\ba
\dfrac{{\cal L}_{G}^{(2)}}{a^3}
&=& - \calR \dfrac{\nabla^2}{a^2} \calR - 3 \dot \calR^2
-18 H \calR \dot \calR + 6 H \alpha \dot \calR + 9 H^2 \alpha \calR
 -2 H \alpha \dfrac{\nabla^2}{a^2} \psi - 3 H^2 \alpha^2
 -\dfrac{27}{2} H^2 \calR^2
\\ \nonumber
&+& 2 \dot\calR \dfrac{\nabla^2}{a^2} \psi -2\alpha \dfrac{\nabla^2}{a^2}\calR\,.
\ea
Eliminating the lagrange multipliers and the other fields in favor of $\cal R$,
we obtain the Lagrangian for $\calR$ as
\ba
\label{R-action}
\dfrac{{\cal L}_\calR^{(2)}}{a^3}
= \mPl^2 \left[ \dfrac{\epsilon}{c_s^2} \dot \calR^2
- \dfrac{\epsilon}{a^2} (\partial \calR)^2  \right]\,,
\ea
where  $\epsilon \equiv - \dot H/ H^2$ is the slow-roll parameter
as defined before. Note, however, that the action~(\ref{R-action}) is
obtained without any slow-roll assumptions.
Also one can check that the above quadratic action results in the same
linear equation for $\cal R$ as given in Eq.~(\ref{R-eq}).

Now let us quantize the system.
Changing the time variable $t$ to the conformal time $\tau$,
the quadratic action~(\ref{R-action}) becomes
\ba
S=\dfrac{1}{2} \int d^3 x d \tau z^2
\left[ \calR'^2 - c_s^2 (\nabla \calR)^2 \right]\,,
\ea
where the prime denotes a derivative with respect to the conformal time and
\ba
z^2 \equiv \dfrac{2\epsilon a^2}{c_s^2} \mPl^2  \, .
\ea
The momentum conjugate to the field $\calR$ is
\ba
\Pi_\calR\equiv\frac{\delta S}{\delta\calR'}=z^2 \calR'  \, .
\ea
They satisfy the canonical commutation relation,
\ba
\left[ \calR(\vec{x},\tau) \, , \, \Pi_\calR(\vec{y},\tau)  \right]
 = i \delta^3(\vec x -\vec y) \, .
\ea
The quantized field can be expressed in the Fock
representation,
\ba
\calR({\bf x},\tau) = \int \dfrac{d^3 k}{(2 \pi)^3}
\left[\calR_k(\tau) a_{\bf k} e^{i {\bf k.x}}
 + \calR^*_k(\tau) a^\dagger_{\bf k} e^{-i {\bf k.x}} \right]  \, ,
\ea
where $\calR_k$ is a positive frequency mode function that
satisfies the equation of motion,
\begin{eqnarray}
(z^2\calR')'+c_s^2k^2z^2\calR=0\,,
\end{eqnarray}
and the normalization condition,
\ba
\label{normalization}
\calR_k \calR'^*_k - \calR^*_k \calR'_k = \dfrac{i}{z^2}  \, .
\ea
The annihilation and creation operators, $a_{\bf k}$
and $a^\dag_{\bf k}$, satisfy
\ba
\left[a_{\bf k} \, , \, a^\dagger_{\bf k'}\right]
 =  (2 \pi)^3 \delta^3 ({\bf k-k'}) \,.
\ea
Assuming that $\calR_k$ should approach a conventional positive frequencty
function at high frequencies, $\calR_k\propto e^{-ic_s\,k\tau}$ for
$\tau\to-\infty$, the solution is uniquely determined as
\ba
 \calR_k = C_k x^\nu H_\nu^{(1)}(x) \, ,
\ea
where $H_\nu^{(1)}$ is the Hankel function of the first kind,
\ba
\label{nu-eq}
x=-c_s k (\tau - \tau_e -{\cal H}_e^{-1})\,,
\qquad \nu= \dfrac{3+\eta}{2}  \,,
\ea
and
\ba
\label{C}
\vert C_k \vert^2 = \dfrac{\pi c_s}{8 k \epsilon_i a_i^2 \mPl^2} x_i^{1-2\nu}  \,.
\ea
Here again the subscript $i$ denotes an initial/reference time $\tau=\tau_i$.
One might suspect that the absolute value of $C_k$ would depend on the
choice of the initial time $\tau_i$. However, for a nearly constant $\eta$,
one can show that it is independent of $\tau_i$ because one has
$\epsilon\,a^2\propto a^{\eta+2}$ and $x^{1-2\nu}\propto a^{2\nu-1}=a^{\eta+2}$.

One of the important properties of our model is that the curvature
perturbation is not conserved after horizon crossing. Expanding the
Hankel function at $x\ll1$ gives
\ba
\label{Rsuper}
\calR_k(\tau) \simeq -  C_k \dfrac{i 2^{-\nu} e^{-i\pi \nu}}{\pi}
 \Gamma(\vert \nu \vert) x(\tau)^{2 \nu}\,.
\ea
As a result, the final curvature perturbation at the end of ultra slow-roll
inflation $\tau=\tau_e$  is given by
\ba
\label{Rsuperf}
\calR_k(\tau_e) \simeq -  C_k \dfrac{i 2^{-\nu} e^{-i\pi \nu}}{\pi}
\Gamma(\vert \nu \vert)  \left( \dfrac{c_s k}{\calH_e}  \right)^{2 \nu}\,.
\ea

The power spectrum of curvature perturbation at the end of
ultra slow-roll inflation
is given by
\ba
{\cal P_R}= \dfrac{k^3}{2 \pi^2} \vert \calR_k(\tau_e) \vert^2\,.
\ea
By using Eq.~\eqref{C} the above reduces to
\ba
{\cal P_R} \simeq \dfrac{\Gamma(\vert \nu \vert)^2 }{\pi^3 2^{2\nu+4} }
\left(\dfrac{H_e}{\mPl} \right)^2 \dfrac{1}{c_s\epsilon_e}
\left( \dfrac{c_sk}{H_e a_e} \right)^{3+2\nu} \,,
\ea
which, using the approximation $\eta \simeq -3(1+c_s^2)$,
further reduces to
 \ba
{\cal P_R} \simeq \dfrac{\Gamma(3c_s^2/2)^2 }{ \pi^3 2^{4-3c_s^2} }
\left(\dfrac{H_e}{\mPl} \right)^2  \dfrac{1}{c_s \epsilon_e}
\left( \dfrac{c_s k}{H_e a_e} \right)^{3(1-c_s^2)}  \, .
\ea

The spectral index is easily read off as
\ba
\label{ns-eq}
n_s-1 \simeq 3+2\nu \simeq 3(1-c_s^2)\,.
\ea
Interestingly, the sound speed explicilty appears in the spectral index in
this model, in contrast to the standard inflationary scenarios in
which only  $\dot c_s$ plays a role in the spectral index.
In order to have a scale-invariant perturbations we require $c_s=1$.
The amplitude of the spectrum in this case is given by
\begin{eqnarray}
{\cal P_R} = \frac{H^2}{8\pi^2{\mPl}^2}\dfrac{1}{\epsilon_e}\,.
\label{PRamplitude}
\end{eqnarray}

A red tilted power spectrum can be achieved by a slightly superluminal
sound speed. With $c_s=1$, from Eq. (\ref{eta-eq}) we obtain
$\eta \simeq -6$ and from Eq. (\ref{nu-eq}) $\nu \simeq -3/2$. This yields
$\epsilon \propto a^{-6}$ as mentioned before.
Of course, recent cosmological observations by WMAP and PLANCK strongly favor
a red-tilted power spectrum \cite{Ade:2013uln}. We see that in our model
a subluminal sound speed implies $n_s >1$. This is a direct consequence of
the starting assumption of our considering an isentropic fluid.
To obtain a red spectral index for a subluminal sound speed,
perhaps one should consider a more general, non-isentropic fluid.

As for the tensor to scalar ratio, since the tensor spectrum is exactly
the same as the standard case,
\begin{eqnarray}
{\cal P}_T=\frac{2H^2}{\pi^2\mPl^2}\,,
\end{eqnarray}
one finds
\ba
r=\frac{{\cal P}_T}{\cal P_R} =16 \epsilon_e\,.
\ea
Since $\epsilon$ decreases exponentially during ultra slow-roll inflation,
we conclude that the amplitude of the tensor perturbation is
exponentially suppressed in this model.

The above simple model is not complete by itself, since there is no
mechanism to terminate inflation. In principle, one can match the
non-attractor phase of inflation to an attractor phase
of conventional slow-roll inflation or of a hot Friedmann stage
at which $\epsilon$ is not decaying exponentially. At such a second stage,
${\cal R}$ becomes frozen on super-horizon scales as usual.
This implies that one can read off the final value of ${\cal R}$
by computing its value at $\tau=\tau_e$ when the transition from the
non-attractor phase to an attractor phase starts.
This picture was employed in the context of a single scalar
field theory in \cite{Namjoo:2012aa}.
The second phase of inflation is necessary also because the non-attractor inflationary phase we considered here cannot last long enough to solve the horizon problem. Because the slow-roll parameter is decreasing exponentially with time, to get ${\cal P_R} \sim 6\times 10^{-9}$, we need a low-scale $H$ \cite{Chen:2013aj}. For example, if we assume the lower-bound reheating energy to be $\sim 1 ~{\rm GeV}$, we have $\epsilon_{\rm min} \sim 10^{-66}$. This means that the upper bound of the inflationary efold for this non-attractor phase is $25$.

\subsection{Cubic Action and non-Gaussianity}

Here we expand the action to third order which will be suitable to calculate
the bispectrum. Starting with the action given in Eq.~(\ref{S-general}), one has
\ba
S_3 = \int d^4 x \left[ \mPl^2 {\cal L}_{G}^{(3)}
+ a^3 \left( L_m^{(3)} + (\alpha+3 \calR) L_m^{(2)}
 + (3 \alpha \calR +\dfrac{9\calR^2 }{2}) L_m^{(1)}
 + \dfrac{9}{2}( \calR^2 \alpha +  \calR^3) L_m^{(0)}\right) \right]
\ea
where ${\cal L}_{G}^{(3)}$ represents the cubic order gravitational
Lagrangian density, while ${ L}_m^{(i)}$ stands for the
$i$-th order matter Lagrangian.

With the expansion,
\ba
E (\x, t) &=& \rho(\x,t) \left (1+e(\x,t) \right)
\simeq  E + (E+P) \dfrac{\delta \rho}{\rho}
+ c_s^2 \epsilon E \dfrac{\delta \rho^2}{3 \rho^2} +
\dfrac{c_s^2 \epsilon E}{27 \rho^3} (-2s + 2\epsilon -\eta -6 ) \delta \rho^3\,,
\\
\rho(\x,t) U^0(\x,t) &\simeq &\rho -\alpha \rho +\delta\rho
- \alpha \delta \rho + \alpha^2 \rho + \alpha^2 \delta \rho -\alpha^3 \rho\,,
\\
g_{\mu \nu} U^\mu U^\nu &\simeq & -1\,,
\ea\\
one can check that
\ba
{ L}_m^{(3)} = \dfrac{\dot \calR^3}{H^3} E \epsilon
\left[ \frac{\left(2\epsilon-2 s-\eta-6 \right) }{27 c_s^4}
- \dfrac{2}{3} (1+\dfrac{1}{c_s^2}) \right]\,,
\ea
where we have introduced
\ba
s \equiv \frac{ \dot c_s}{H c_s } \, .
\ea
(Not to be confused with the entropy density.)

Using the constraint equations to remove non-dynamical variables,
the full cubic action from the matter sector is
\ba
\label{cubic-fluid}
 (N \sqrt{h} { L}_m)\vert_{(3)} = - \left(2 \tilde \lambda
+ \tilde \Sigma \right) \dfrac{\dot \calR^3}{H^3}
+ 3 \tilde \Sigma \dfrac{\dot \calR^2 \calR }{H^2}
- \dfrac{9 \calR^2 \dot \calR}{2H} E + \dfrac{9}{2} P \calR^3\,,
\ea
where
\ba
\tilde{\Sigma} &\equiv& \dfrac{\mPl^2 H^2 \epsilon}{c_s^2}\,,
\\
\tilde{\lambda} &\equiv &
\dfrac{\tilde{\Sigma} }{18 c_s^2} (\eta +6 +2(s-\epsilon))
= \dfrac{\tilde{\Sigma} }{6 c_s^2} \left(\frac{2 s}{3} - c_s^2 + 1 \right)\,.
\ea

As demonstrated in Appendix \ref{appendix}, similar to \cite{Garriga:1999vw} and \cite{Arroja:2010wy},
one can check that the above cubic matter Lagrangian is equivalent to
that for the theory of a scalar field with the action ${\cal L}_m =P(X)$
where $X= -g^{\mu \nu} X_\mu X_\nu/2$,
$\tilde \Sigma =X P_X+ 2 X^2 P_{XX}$ and
$\tilde\lambda = X^2 P_{XX} + \dfrac{2}{3} X^3 P_{XXX}$.
Since the gravitational part of the action is the same by construction,
this conclusion enables us to cast the cubic action for our model to the
well-studied cubic action for a general $P(X, \phi)$ theory for
k-inflation~\cite{ArmendarizPicon:1999rj, Garriga:1999vw} or DBI
inflation~\cite{Alishahiha:2004eh} with \cite{Seery:2005wm, Chen:2006nt}
\begin{eqnarray} \label{action3}
S_3&=&\int dt d^3x\{
-a^3 (\tilde \Sigma(1-\frac{1}{c_s^2})+2 \tilde \lambda)\frac{\dot{\calR}^3}{H^3}
+\frac{a^3\epsilon}{c_s^4}(\epsilon-3+3c_s^2)\calR\dot{\calR}^2
\nonumber \\ &+&
\frac{a\epsilon}{c_s^2}(\epsilon-2s+1-c_s^2)\calR(\partial\calR)^2-
2a \frac{\epsilon}{c_s^2}\dot{\calR}(\partial
\calR)(\partial \chi) \nonumber \\ &+&
\frac{a^3\epsilon}{2c_s^2}\frac{d}{dt}(\frac{\eta}{c_s^2})\calR^2\dot{\calR}
+\frac{\epsilon}{2a}(\partial\calR)(\partial
\chi) \partial^2 \chi +\frac{\epsilon}{4a}(\partial^2\calR)(\partial
\chi)^2+ 2 f(\calR)\frac{\delta L}{\delta \calR}|_1 \} ~,
\end{eqnarray}
where the field $\chi$ is defined by
\ba
\partial^2 \chi =a^2 \frac{\epsilon}{c_s^2} \dot \calR\, ,
\ea
and $f(\calR)$ and $\delta L/\delta \calR|_1$, respectively, by
\begin{eqnarray}
\label{redefinition}
f(\calR)&=&\frac{\eta}{4c_s^2}\calR^2+\frac{1}{c_s^2H}\calR\dot{\calR}+
\frac{1}{4a^2H^2}[-(\partial\calR)(\partial\calR)
+\partial^{-2}(\partial_i\partial_j(\partial_i\calR\partial_j\calR))]
\nonumber \\
&+&
\frac{1}{2a^2H}[(\partial\calR)(\partial\chi)
-\partial^{-2}(\partial_i\partial_j(\partial_i\calR\partial_j\chi))]\,,
\end{eqnarray}
and
\begin{eqnarray}
\frac{\delta
L}{\delta\calR}\mid_1 &=& a
\left( \frac{d\partial^2\chi}{dt}+H\partial^2\chi
-\epsilon\partial^2\calR \right)\,.
\end{eqnarray}

So far our analysis of the cubic action was general and no assumption on the value of
$c_s$ has been made. However, from our power spectrum analysis,
Eq.~(\ref{ns-eq}), we see that to obtain a scale-invariant power
spectrum we need $c_s =1$. Therefore, from now on we concentrate on the case
$c_s=1$. In this limit, all the interaction terms in the cubic action
becomes small except for the last term involving $f(\calR)$.
It is known that this last term can be eliminated by
the field redefinition $\calR \rightarrow \calR_n+f(\calR_n)$.
This means that the leading contribution to non-Gaussianity comes only
from the field redefinition.
As emphasized in \cite{Namjoo:2012aa} both of the first two terms in
$f(\calR)$ in Eq.~(\ref{redefinition}) contribute to non-Gaussianity.
This is in contrast to the usual attractor situation in which $\dot \calR $
vanishes on the super-horizon scales
and only the first term in $f(\calR)$ contributes to non-Gaussianity.

Following the same steps as in \cite{Namjoo:2012aa}, the amplitude of
local type non-Gaussianity, $f_{NL}$, defined in the squeezed limit,
$k_1\ll k_2=k_3$, as
\ba
\langle \calR_{\bf k_1} \calR_{\bf k_2} \calR_{\bf k_3}\rangle \simeq
(2\pi)^3 \delta^3( \sum_i {\bf k_i} )\, \frac{12}{5}f_{NL}P_{k_1}P_{k_3}\,,
\ea
is obtained to be
\ba
\label{fNL}
f_{NL}= - \frac{5}{4} \left( \eta +4 \right)
=\dfrac{5}{2}  \, .
\ea
This value of $f_{NL}$ is consistent with the recent Planck constraints
on primordial non-Gaussianity \cite{Ade:2013ydc}.


\section{A model}
\label{Model}
Here we present a single field model which shows the behavior similar to
 what we pointed out in the previous sections. Consider a canonically
normalized field, so $c_s=1$, with the potential,
\ba
\label{V}
V(\phi)=
\begin{cases}
V_0 \qquad  &\mbox{for}~\phi<\phi_c\,,
\\
 V_1 (\phi) \qquad&\mbox{for}~\phi>\phi_c\,.
\end{cases}
\ea
During the first stage, the system approaches rapidly towards a de Sitter
universe since $\epsilon \propto a^{-6}$.
This model was originally studied in \cite{Kinney:2005vj} as
``ultra slow-roll'' (USR) and was  further studied in \cite{Namjoo:2012aa} as
a toy single field model which can produce non-negligible local non-Gaussianity.
During this phase, the curvature perturbation is not frozen on super-horizon
scales, exhibitin the non-attractor nature of the system.
As studied in \cite{Namjoo:2012aa}, the background dynamics
during the non-attractor phase is
\ba
\ddot \phi + 3 H \dot \phi =0\,,
\quad  3 M_P^2 H^2 = \frac{\dot \phi^2}{2} + V_0 \simeq V_0\,.
\ea
Thus $\dot \phi \propto a^{-3}$ and hence
\ba
\epsilon \propto a^{-6} \,,\quad \eta \simeq -6 \, .
\ea
The power spectrum and bispectrum were computed in \cite{Namjoo:2012aa},
and the local-type non-Gaussianity with $f_{NL}=5/2$ was obtained.

It is instructive to look at the bispectrum in the squeezed limit
using the $\delta N$ method. One has
\begin{eqnarray}
N(\phi,\dot\phi)=\frac{1}{3}\ln
\left[\frac{\dot\phi}{\dot\phi+3H(\phi-\phi_c)}\right]\,,
\label{Nform}
\end{eqnarray}
where $N$ is the number of $e$-folds counted backward from
the end of ultra slow-roll inflation at which $\phi=\phi_c$
(not to be confused with the lapse function).
It is important to note that $N$ is a function not only of $\phi$
but also of $\dot\phi$, in contrast to the conventional slow-roll inflation
for which $\dot\phi$ is not independent but a function of $\phi$.
Taking the variations of $\phi$ and $\dot\phi$ yields
\begin{eqnarray}
\delta N=N(\phi+\delta\phi,\dot\phi+\delta\dot\phi)-N(\phi,\dot\phi)\,.
\label{deltaN}
\end{eqnarray}
On super-horizon scales, $\delta \phi$ follows the evolution of
background $\phi$, and one can check that $\delta \dot \phi\simeq 0$
on super-horizon scales. As a result
\begin{eqnarray}
\delta N
&\simeq&\frac{\partial N}{\partial\phi}\delta\phi
+\frac{1}{2}\frac{\partial^2 N}{\partial\phi^2}\delta\phi^2
\nonumber\\
&=&-\frac{H}{\dot\phi+3H(\phi-\phi_c)}\delta\phi
+\frac{3H^2}{2\Bigl(\dot\phi+3H(\phi-\phi_c)\Bigr)^2}\delta\phi^2
\,.
\end{eqnarray}
This automatically yields $f_{NL}=5/2$ in agreement  with the result obtained
from the in-in formalism.

As mentioned before, inflation never ends unless there is a
mechanism to terminate the non-attractor phase.
In the current example, we have introduced a non-trivial potential
for $\phi>\phi_c$. At the second phase, inflation proceeds as in
the conventional slow-roll inflation and $\cal R$ freezes out
on super-horizon scales. Therefore, the physical
parameters such as $f_{NL}$ and $n_s$ can be read off by calculating
these quantities at $\tau=\tau_c$ when the non-attractor phase is matched
to the attractor phase. \\

In summary, in this work we have presented a fluid description of inflation.
To be specific, we have considered the action of a  single barotropic perfect
fluid with appropriate Lagrange multipliers. After eliminating the Lagrange
multipliers and the other non-dynamical variables we have obtained
the quadratic and cubic actions for $\calR$.
We have shown that this barotropic fluid naturally gives rise to
a non-attractor inflationary phase in which $\calR$ is not frozen on
super-horizon scales. An interesting prediction of this model is that
the curvature perturbation power spectrum is scale-invariant
with the value of local type non-Gaussianity given by $f_{NL}=5/2$.
We have also shown that at the level of cosmological perturbation theory
this fluid model is equivalent to a scalar field theory with
the Lagrangian $P(X)$.

The natural question which arises is how one can extend this formalism to
a non-barotropic fluid for which the pressure is not uniquely determined
by the energy density.
This may help to keep $n_s$ as a free parameter to obtain a slightly
red-tilted power spectrum as suggested by the PLANCK data \cite{Ade:2013uln}
without appealing to a superluminal fluid.
However, this may also result in generating entropy perturbations which
are under strong observational constraints by the PLANCK data \cite{Ade:2013uln}. We would like to come back to this issue elsewhere.

Also in this work we have considered a model with constant $c_s$. In principle one may relax this assumption and consider the case in which
 $c_s$ is time-dependent. As a result, this will add the new contribution $\dot{c}_s/c_s$ (in the limit where $c_s$ is changing slowly with time) into $n_s$. It is an interesting question to see if this can help to obtain a red-tilted power spectrum.\\

\begin{acknowledgements}
 We would like to thank Nima Khosravi,
Javad Taghizadeh Firouzjaee, Alberto Nicolis and Jonathan White for useful discussions.
This work was supported in part by the JSPS Grant-in-Aid for Scientific
Research (A) No.~21244033.
\end{acknowledgements}

\appendix

\section{The action for $P(X, \phi)$ theory}
\label{appendix}

In this appendix we prove the equivalence between the perturbation theory
in our isentropic fluid and a scalar field theory with the matter action,
\ba
L_M =  P(X,\phi)  \, ,
\quad
X \equiv -\frac{1}{2} g^{\mu \nu} \partial_\mu \phi \partial_\nu \phi\,,
\ea
similar to k-inflation models \cite{ArmendarizPicon:1999rj, Garriga:1999vw}.
This equivalence will be used to map the bispectrum in our model to that of
a well-studied $P(X, \phi)$ theory, such as in  \cite{Chen:2006nt}.

Our aim here is to expand the matter Lagrangian up to third order of
perturbations. It is convenient to adopt the comoving gauge
in which $\delta \phi=0$ and
\ba
\delta X = \frac{\delta g^{00}}{g^{00}} X \simeq
\left( - 2\alpha + 3 \alpha^2 - 4 \alpha^3 \right)X\,.
\ea
Noting  that $\alpha = \dot \calR/H$,  up to third order in comoving
gauge we have
\ba
\sqrt{h} &\simeq& a^3 (1+3 \calR + \dfrac{9}{2} \calR^2
+ \dfrac{9}{2} \calR^3)\,,
\\
N &\simeq& 1+ \dfrac{\dot \calR }{H}\,,
\\
P(X,\phi) &\simeq& P-X P_X  \left(2 \dfrac{\dot \calR}{H}
 - 3 \dfrac{\dot \calR^2}{H^2}+4 \dfrac{\dot \calR^3}{H^3} \right)
+2 X^2 P_{,XX} \left( \dfrac{\dot \calR^2}{H^2}
-3 \dfrac{\dot \calR^3}{H^3}\right)
-\dfrac{4}{3} X^3 P_{,XXX} \dfrac{\dot \calR^3}{H^3}\,.
\ea
Gathering all cubic order terms we obtain
\ba
\label{P-cubic}
 (N \sqrt{h} { L}_M)\vert_{(3)}
 = - \left(2  \lambda +  \Sigma\right) \dfrac{\dot \calR^3}{H^3}
 + 3 \Sigma \dfrac{\dot \calR^2 \calR }{H^2}
- \dfrac{9 \calR^2 \dot \calR}{2H} E + \dfrac{9}{2} \calR^3 P\,,
\ea
where $E= 2 X P_X - P$ is the total energy density
that appears in the Friedmann equation, $3 M_P^2 H^2 = E$.

Comparison between Eq. (\ref{P-cubic}) and Eq. (\ref{cubic-fluid})
demonstrates the equivalence between the above theory and the matter
sector of our fluid theory with the identifications
$\tilde \Sigma  \leftrightarrow \Sigma$
and $\tilde \lambda \leftrightarrow \lambda$.

{}

\end{document}